\documentclass[%
reprint,
superscriptaddress,
 prl,
]{revtex4-2}

\usepackage{graphicx}% Include figure files
\usepackage{dcolumn}% Align table columns on decimal point
\usepackage{bm}% bold math
\usepackage{hyperref}% add hypertext capabilities
\usepackage[mathlines]{lineno}% Enable numbering of text and display math
\usepackage[T1]{fontenc}
\usepackage{amsmath}
\usepackage{natbib}

\usepackage{xcolor}

\begin{document}

\preprint{APS/123-QED}

\title{{\color{black} Transient logic operations} in acoustics through dynamic modulation}

\author{Zhao-xian Chen}

    \affiliation{College of Engineering and Applied Sciences, Collaborative Innovation Center of Advanced Microstructures, and National Laboratory of Solid State Microstructures, Nanjing University, Nanjing, 210023, China}
   
\author{Ling-ling Ma}

    \affiliation{College of Engineering and Applied Sciences, Collaborative Innovation Center of Advanced Microstructures, and National Laboratory of Solid State Microstructures, Nanjing University, Nanjing, 210023, China}
   
\author{Shi-jun Ge}

    \affiliation{College of Engineering and Applied Sciences, Collaborative Innovation Center of Advanced Microstructures, and National Laboratory of Solid State Microstructures, Nanjing University, Nanjing, 210023, China}
   
\author{Ze-Guo Chen}
\email{zeguoc@nju.edu.cn}

    \affiliation{School of Materials Science and Intelligent Engineering, Nanjing University, Suzhou, 215163, China.}

\author{Yan-qing Lu}
    \email{yqlu@nju.edu.cn}
    \affiliation{College of Engineering and Applied Sciences, Collaborative Innovation Center of Advanced Microstructures, and National Laboratory of Solid State Microstructures, Nanjing University, Nanjing, 210023, China}
\date{\today}% It is always \today, today,

\begin{abstract}
{\color{black}In quantum logic operations, information is carried by the wavefunction rather than the energy distribution. Therefore, the relative phase is essential. Abelian and non-Abelian phases can be emulated in classical waves using passive coupled waveguides with geometric modulation.} However, the dynamic phases interference induced by waveguide structure variation is inevitable. {\color{black}To overcome the challenges, we introduce an electroacoustic coupled system that enables the precise control of phase distribution through dynamic modulation of hopping.} Such effective hopping is electronically controlled and is utilized to construct various paths in parameter space. {\color{black}These paths lead to state evolution with matrix-valued geometric phases, which correspond to logic operations.} We report experimental realizations of several logic operations, including $Y$ gate, $Z$ gate, Hadamard gate and non-Abelian braiding. Our work introduces a temporal process to manipulate transient modes in a compact structure, providing a versatile experimental testbed for exploring other logic gates and exotic topological phenomena.
\end{abstract}

\maketitle
Adiabatic evolution of states encodes the global topological properties. Topology renders the system stable, associated with robust behavior and observable quantity. Remarkable examples include the geometric phase \cite{Berry1984,Berry1990,wilczek1984appearance}; quantized charge transport in the Thouless pumping \cite{Thouless1983}; nontrivial band topology and corresponding boundary effects in the Floquet topological insulators \cite{PhysRevB.79.081406,lindner2011floquet}. Replicating these achievements for other wave realms beyond electronics offers opportunities to explore unconventional wave phenomena. The coupled waveguide arrays system (CWAS) provides an elegant platform to bridge the Schrödinger equation and the classical waves, and is of practical importance to realize the state evolution. Being rooted in these analogues, the feature of topological protection is transferred to corresponding novel wave phenomena, including braiding classical states \cite{noh2020braiding,chen2022classical,zhang2022non,you2022observation}, robust edge states in Floquet topological photonic or acoustic waveguides \cite{khanikaev2013photonic,peng2019chirality,pyrialakos2022bimorphic,zhu2022time}, transportation of edge states to arbitrary boundaries \cite{kraus2012topological,zilberberg2018photonic,rosa2019edge,shen2019one,chen2021landau,chen2021creating,XU20221950}, mode conversion by encircling the exceptional point \cite{doppler2016dynamically,PhysRevX.8.021066,PhysRevLett.125.187403} and so on. 

Constrained by the state lifetime and modulation rate, considerable technological challenges persist in driving the state dynamically \cite{xu2016topological,cheng2020experimental}. Thus, the current experimental work to investigate state evolution in classical wave system rely on employing a space dimension, i.e., the propagation dimension $z$, to replace the role of time $t$. A continuous variation of the geometric parameters in the CWAS can implement the temporal modulation. Though such a strategy based on solving the stationary equation is liberated from state lifetime constraints and is sufficient for many applications, there is significant merit in elevating space modulations to temporal ones. For instance, we note a dynamic phase difference when we reinspect its application in braiding waves \cite{chen2022classical}. The braiding manifests as state exchange with certain phases, and its non-Abelian generalization provides the necessary background on which holonomic computation plays out \cite{zanardi1999holonomic}. Using waves to realize braiding relies on creating strictly degenerate states to mimic identical particles \cite{boross2019poor}. However, the degeneracy is protected by the chiral-symmetry, which is not intrinsic and generally be fragile in the CWAS \cite{PhysRevApplied.14.024023}. {\color{black}The degeneracy requires precise structures; otherwise, the final states would acquire different dynamic phases, which cause error in the final result.} 

{\color{black}Manipulating transient acoustic modes in uniform cavities offer a promising way to overcome these challenges and achieve accurate results.} In this letter, we propose an electroacoustic coupled system that supports transient braiding through dynamic modulation. In contrast to CWAS, wherein the coupling is contributed by additional waveguides or evanescent waves, the effective coupling in our system can be tailored dynamically. By carefully balancing the state lifetime, the virtual hopping strength and the duration of the modulation, we can realize arbitrary $SO(n)$ operators, one of which is the foundation of braiding. {\color{black}Unlike previous methods using CWAS that require specific structures for different operators, our work demonstrates the efficient manipulation of transient states to realize them in a compact and versatile structure.} The methods are universal to inspire diverse wave manipulations through external field-controlled dynamic modulation.  

\begin{figure*}
\begin{center}
\includegraphics[width=17.6cm]{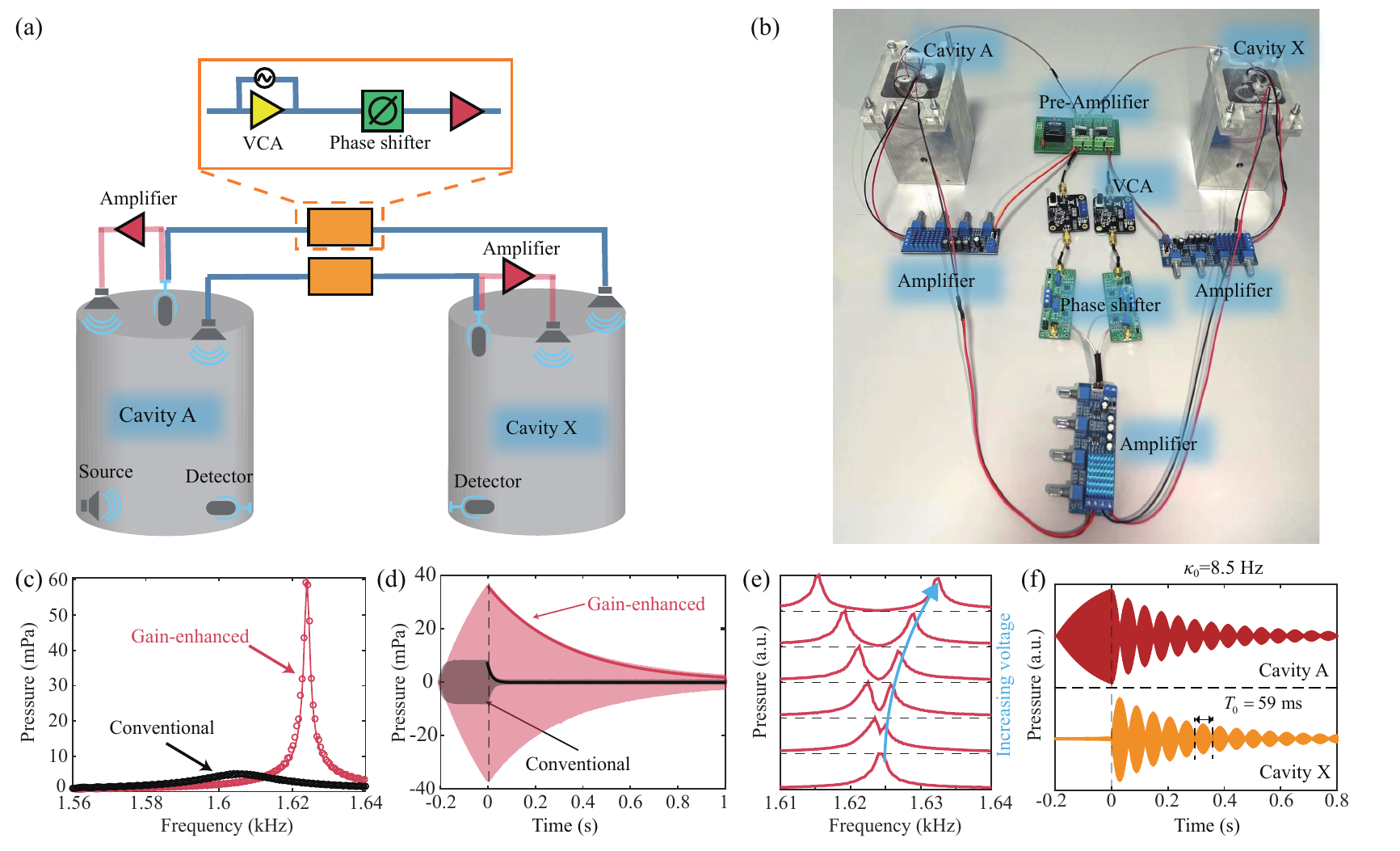}
\end{center}
\caption{\label{fig:1}An electroacoustic coupling platform with dynamic modulation. (a) Schematic of the coupled acoustic cavities to realize the transient evolution of the state. The pink line connected components contribute to a gain-enhanced long lifetime state, and the blue line components induce an effective time-varying coupling. (b) A photograph of the experimental setup shown in (a). (c) The measured spectra for the conventional (black) cavity and gain-enhanced (pink) cavity. (d) Microphone reading of the resonant mode excitation and damping. (e) The response spectrum shows an effective coupling as a function of the control voltage of VCA. (f) Microphone reading inside the two cavities shows oscillating behavior. We excite the cavity mode of frequency 1624 Hz at cavity A during $t$<0 and switch off the excitation during $t$>0, then an effective coupling $\kappa _0=8.5$Hz is switched on.}
\end{figure*}

\emph{Implementation of instantaneous acoustic state evolution.} We begin with a two-state system containing two coupled cavities A and X, having the same first-order resonant frequency $f_0$. The Hamiltonian of the system can be written as \cite{supply}
 \begin{equation}
{\color{black}H_0 = 2\pi\left( {\begin{array}{*{20}{c}}
{{f_0} - i{\Gamma _0} + i{\Gamma _1}}&{\kappa \left( t \right)}\\
{\kappa \left( t \right)}&{{f_0} - i{\Gamma _0} + i{\Gamma _1}}
\end{array}} \right)}
,
\end{equation}
where $\Gamma _0$ denotes the intrinsic loss of each cavity, $\Gamma _1$ indicates additional gain and $\kappa\left( t \right)$ denotes the time-dependent hopping between the two cavities. As depicted in Fig.~\ref{fig:1}(a), here the two acoustic cavities are coupled with external circuits together with the microphones and louder speakers \cite{zhang2021acoustic, chen2022sound}. To observe the transient state evolution, the damping of the system should be pretty small. We use coherent acoustic sources, shown in pink line connected components, to generate a gain response that balances the intrinsic loss. This strategy has been widely used to realize acoustic parity-time symmetric systems with balanced gain and loss \cite{fleury2015invisible,liu2022experimental}. We go a step further to realize the mutual and time-varying coupling by an additional setup. It consists of microphones, voltage control amplifiers (VCAs), phase shifters and loudspeakers. Specifically, the setup amplifies the signal detected from cavity A(X) and couples it to cavity X(A). {\color{black}The electrically-controlled VCA can amplify the sinusoidal signal $sin(\omega t)$ to $V_csin(\omega t)$, in which $V_c$ is proportional to the strength of the mutual coupling $\kappa$, implying an efficient strategy to control the effective hopping between two cavities.} To compensate for the phase introduced by the VCA, we add a phase shifter.

Experimentally, as shown in Fig.~\ref{fig:1}(b), the acoustic cavities are machined precisely from stainless-steel block with depth $h=10$ cm \cite{supply}. Sealing with the rigid acrylic board, the cavity supports the first-order mode eigenfrequency at $f_0=1606$ Hz. Small ports with a diameter of 4 mm are opened for the external detector and speaker, which also introduce radiation loss contained in $\Gamma _0$. In the absence of external gain and coupling, these two modes are essentially uncoupled and have a relatively large attenuation rate $\Gamma _0=8$ Hz. {\color{black}In contrast, for the gain-enhanced cavity, the attenuation rate is reduced to $\Gamma _0-\Gamma _1=0.6$ Hz, which extends the cavity modes’ lifetime. To prevent self-excitation caused by the imaginary part of the effective hopping, the attenuation rate cannot be infinitely small. The optimized value is conducive to the system's stability when relatively large coupling is introduced.} These parameters are extracted from the measured spectra in Fig.~\ref{fig:1}(c), consistent with the results from the measured waveform excited in the resonant frequency shown in Fig.~\ref{fig:1}(d), seeing discussions in \cite{supply}. The resonant frequency is slightly increased to 1624 Hz due to the external effective acoustic load. When the VCAs are in operation, an effective coupling occurs between the two cavities. The coupling strength is controlled by the imposed gate voltage, verified by two resonance peaks in Fig.~\ref{fig:1}(e). In addition, we can flip the sign of the coupling $\kappa$ by electrically controlling the phase shifters \cite{supply}.

To further show the evolution characteristic, we excite cavity A at 1624 Hz for 0.2 s, then immediately switch on the coupling. As shown in Fig.~\ref{fig:1}(f), the acoustic energy oscillates between the two cavities at a periodicity of $59$ ms, consistent with the theoretical results $T _0=1/2\kappa _0$ by solving  $i\frac{{\partial \psi(t) }}{{\partial t}} = H_0\psi(t) $. The excellent agreement provides direct evidence that the Schrödinger type equation governs our system \cite{supply}.  

\begin{figure}
\includegraphics[width=8.3cm]{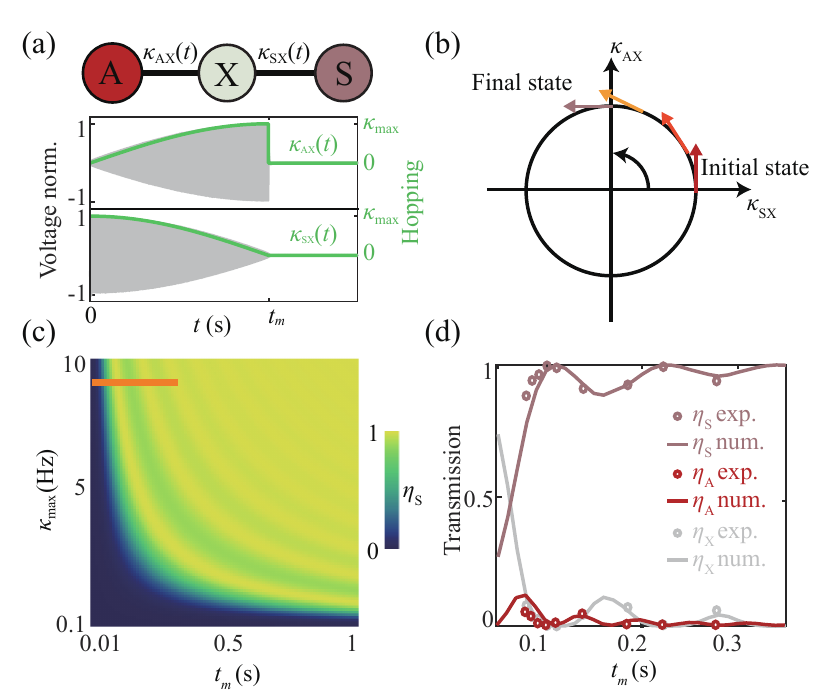}
\caption{\label{fig:2}Experimental verification of transient acoustic energy transfer. {\color{black}(a) The upper panel shows the three-site tight-binding model. The couplings are designed as trigonometric functions (green curves) shown in the lower panel and are implemented by temporally modulating the VCAs to get the targeted outputs (gray domains) when using harmonic signals as the input.} (b) The system sustains one zero mode that evolves as a tangential vector of the parameter circle. (c) The acoustic state transfer ratio $\eta_S$ as a function of the effective hopping strength $\kappa _{\rm{max}}$ and the total evolution time $t _m$. The yellow line denotes the parameter range in (d). (d) Experimental results (dots) of the measured acoustic state transfer as a function of $t _m$ when $\kappa _{\rm{max}}=8.5$ Hz. $\eta_A$ ($\eta_X$) indicates the energy ratio in cavity A (X). Good agreement between the experimental (dots) and numerical results (curves) can be seen.}
\end{figure}

 One notable feature of the VCA is that the control voltage can modulate dynamically. We explore this feature to construct evolution paths wherein the braiding process occurs. To achieve the adiabatic limit, we shall carefully balance the state lifetime, the virtual hopping strength $\kappa _{\rm{max}}$ and the duration of the modulation $t _ m$. We employ a three-site model with dynamic hopping coefficients shown in Fig.~\ref{fig:2}(a) to experimentally study the relations. The corresponding Hamiltonian reads
  \begin{equation}
{H_1(t)} = {\kappa _{\rm{AX}}}(t)\left| \rm{X} \right\rangle \left\langle \rm{A} \right| + {\kappa _{\rm{XS}}}(t)\left| \rm{S} \right\rangle \left\langle \rm{X} \right| + \rm{H.c.}
,
\end{equation}
 where we ignore the onsite energy since all cavities are identical, $\left| \rm{X} \right\rangle$, $\left| \rm{A} \right\rangle$, $\left| \rm{S} \right\rangle$ denote the state localized at corresponding site \cite{boross2019poor}. By setting the hopping as positive and trigonometric functions, the control manifold is isomorphic to a circle defined by the unit hopping vector as shown in Fig.~\ref{fig:2}(b). The chiral-symmetry protected zero mode is $\left| \psi  \right\rangle  = {\kappa _{\rm{XS}}}\left| \rm{A} \right\rangle  - {\kappa _{\rm{AX}}}\left| \rm{S} \right\rangle$, the space of which forms an orthonormal frame bundle on the circle. If the system varies sufficiently slowly (large $t_m$), the state remains to be an eigenstate of $H_1(t)$. A quarter circle control manifold leads to a topological energy transfer from $\left| \rm{A} \right\rangle$ to the state $-\left| \rm{S} \right\rangle$. The negative sign in front of $\left| \rm{S} \right\rangle$ is crucial and is in consequence of the evolution paths in Fig.~\ref{fig:2}(b). Naturally, the $Z$ gate $Z=\left( {\begin{array}{*{20}{c}}
1&{ 0}\\
0&-1
\end{array}} \right)$ can be approached by employing a half circle control manifold combined with an isolated cavity mode \cite{supply}. Even when the process is non-adiabatic, the prefactor -1 exists. We numerically calculate the energy transfer ratio $\eta_S$ as a function of hopping strength $\kappa_{\rm{max}}$ and modulation duration time $t_m$ (see in Fig.~\ref{fig:2}(c)). Though the resulting dynamics oscillate before reaching the adiabatic limit, we note a total energy transfer at $t_m=0.1$ s, which is experimentally feasible. {\color{black}We set $\kappa_{\rm{max}}=8.5$ Hz in this work, and the experimental data in Fig.~\ref{fig:2}(d) agrees well with the numeric results. The agreement implies our system can dynamically manipulate an \emph{instantaneous} state to study the logic operation, in contrast to previous efforts in studying dynamic systems with harmonic modulation to achieve unusual transmission and reflection \cite{koutserimpas2018nonreciprocal,shen2019nonreciprocal,chen2019nonreciprocal,nassar2020nonreciprocity,xu2020physical,williamson2020integrated,wang2022non,wen2022unidirectional}.}    
 
\begin{figure}
\includegraphics[width=8.3cm]{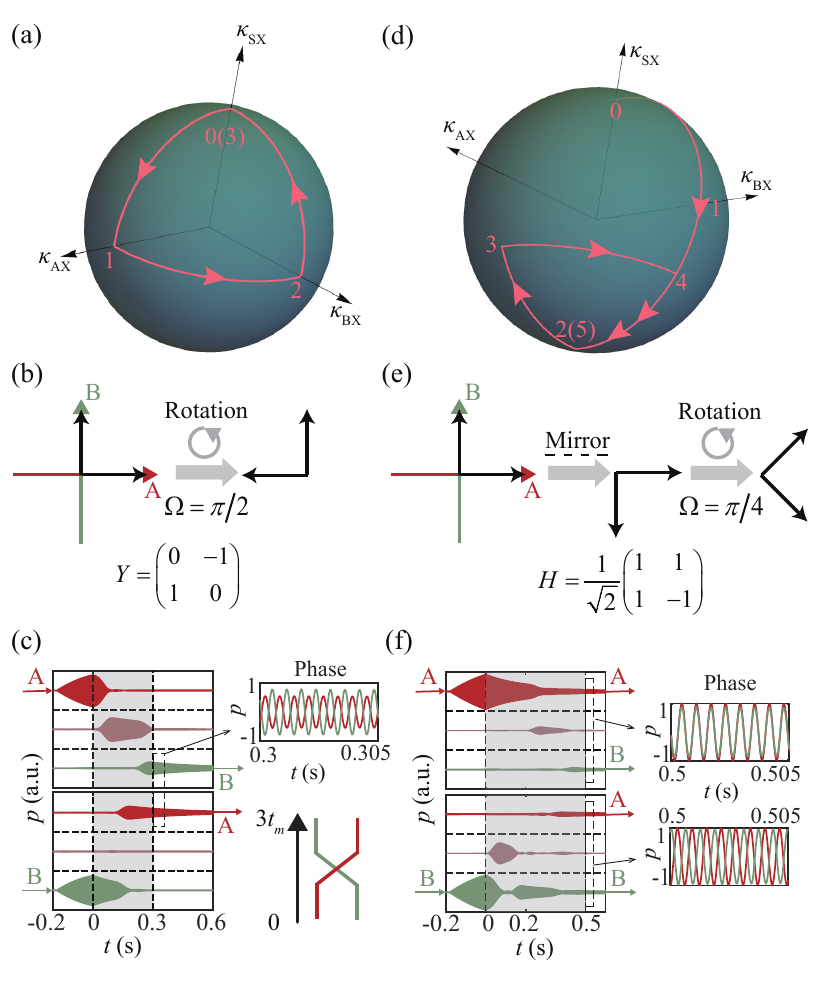}
\caption{\label{fig:3}Experimental realization of transient braiding and the Hadamard gate. (a, d) The control parameters to realize the Y gate (a) and the Hadamard gate (d). Their parallel transports along the paths imply a rotation (b) combined with a reflection operation (e) on a 2D plane. (c, f) Microphone reading shows the acoustic energy transfer behaviors with phase information shown in the right panel, indicating the realization of braiding (c) and Hadamard gate (f). Note that the phase information in (c) is obtained by simultaneous excitation from cavities A and B.}
\end{figure}

\emph{Hadamard gate and non-Abelian braiding of instantaneous acoustic states.} Using the above configuration as the building block, we now extend to a system supporting $n$ degenerate zero states. By varying the control parameters to execute a closed loop in the control manifold $S^n$, the final states undergo a rotation in a $n$D space characterized by an element in $SO(n)$ group \cite{supply}. The group is non-Abelian when $n>2$. If $n=2$, the element can be represented by $U(\Omega)=\left( {\begin{array}{*{20}{c}}
\rm{cos}(\Omega)&{ -\rm{sin}(\Omega)}\\
\rm{sin}(\Omega)&\rm{cos}(\Omega)
\end{array}} \right)$ in which $\Omega$ is the solid angle enclosed by the closed loop in $S^2$. Braiding of two Majorana zero modes results a geometric phase $\left( {\begin{array}{*{20}{c}}
0&{ - 1}\\
1&0
\end{array}} \right)$, which is exactly the Y gate $Y=U(\pi/2)$ manifested as a $\pi/2$ rotation of two degenerate states in 2D space. We plot the control parameters in Fig.~\ref{fig:3}(a), and the states evolution diagram in Fig.~\ref{fig:3}(b) reveals the $Y$ gate operation. This operation, together with its inverse by reversing the swapping sequence, is mapped to the generating set of braid group $B_2$ \cite{chen2022classical,zhang2022non}. {\color{black} The braid operation is sufficient to realize universal logic gates through an exquisite combination \cite{kauffman2004braiding}. In addition, the implemented Z gate expands the scope of realized operations in $SO(2)$ group to operations in $O(2)$ group. For instance, the well-known Hadamard gate, which enables the generation of superposition states, can be approached by a $Z$ gate sequenced by $U(\pi/4)$, i.e., $H=U(\pi/4)\times Z$. The Hadamard gate is widely used in various quantum algorithms. The control parameters in Fig.~\ref{fig:3}(d) can induce the Hadamard gate in our setup. The process corresponds to a reflection operation ($Z$) combining with a rotation operation ($U(\pi/4)$), as shown in Fig.~\ref{fig:3}(e)}. 

To experimentally demonstrate the designed operators, we encode the VCAs to realize the transient hopping modulation and individually excite the state $\left| \rm{A} \right\rangle$ and $\left| \rm{B} \right\rangle$ at the bottom of each cavity for 0.2 s to prepare the initial states \cite{supply}. We document the waveform measured from each cavity in Fig.~\ref{fig:3}(c) and successfully observe the state relocation. Though the loss dampens the state, the relative acoustic energy transfer efficiency exceeds 0.99 after the braiding, where cavity S performs as an intermediary cavity. Moreover, the predicted $\pi$ phase difference is verified by simultaneously exciting the two degenerate states and comparing the waveform after braiding. The characteristic $\pi$ phase of the basic braid operation is illustrated in the right panel of Fig.~\ref{fig:3}(c) and is robust against the chiral-symmetry-preserved perturbations \cite{supply}. Similar experiments are implemented to construct the Hadamard gate. We show its acoustic realization in Fig.~\ref{fig:3}(f). We excite the state A(B) and record the waveform in Fig.~\ref{fig:3}(f), respectively. {\color{black} The output state will be equally distributed in cavity A and B with phase difference 0 ($\pi$) when excited from cavity A (B).} Both intensity and phase information validate our predictions that the Hadamard gate would create an equal superposition state if a basis state is excited.    

\begin{figure}
\includegraphics[width=8.3cm]{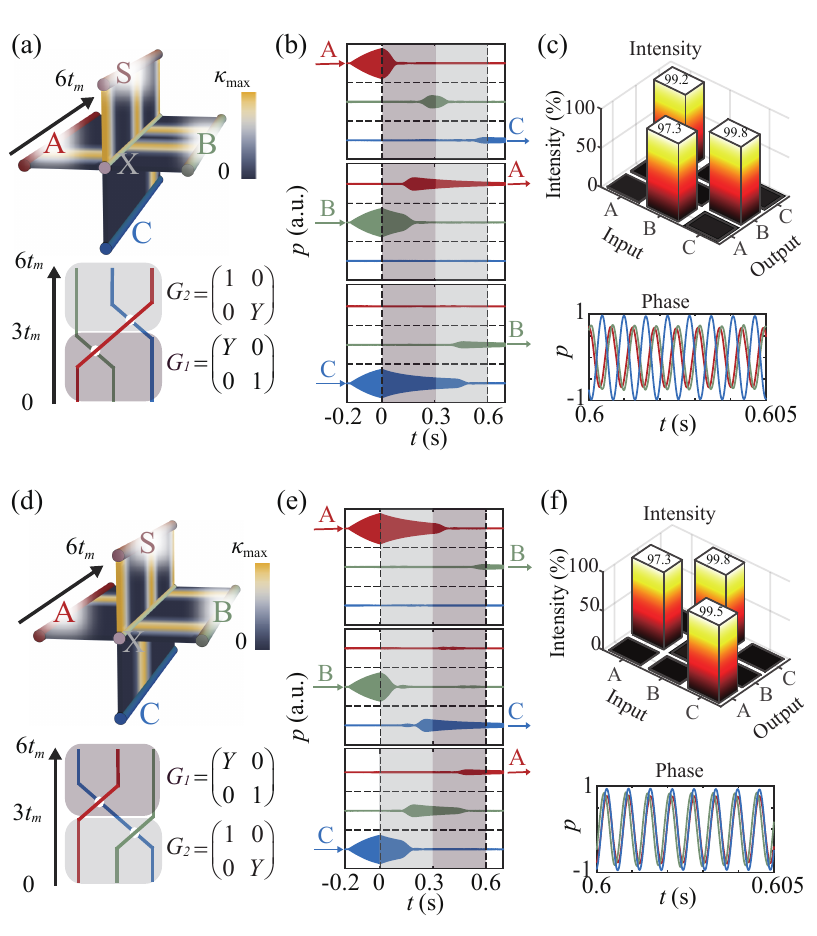}
\caption{\label{fig:4}Experimental realization of transient non-Abelian braiding. A five-site tight-binding model to realize non-Abelian $G_2G_1$ (a) and $G_1G_2$ (d) braiding in which the hopping coefficients are dynamically modulated as the gradient colors. Microphone reading when the system is situated in $G_2G_1$ (b) or $G_1G_2$ (e) configuration. {\color{black}The output distributions and relative phases in (c, f) demonstrate the successful realization of $G_2G_1$  and $G_1G_2$ braiding operations. Specifically, the output state $\left| \rm{C} \right\rangle$ is out of phase compared to output states $\left| \rm{A} \right\rangle$ and $\left| \rm{B} \right\rangle$ after $G_2G_1$ operation (c), while all output states are in phase after $G_1G_2$ operation (f).}} 
\end{figure}

Our system is also extendable for multiple-states evolution. It is straightforward to introduce a cavity C coupled to cavity X to support one more zero mode. Three-state braiding can be captured by matrix product $G_2G_1$ and $G_1G_2$, respectively, where $G_1=\left( {\begin{array}{*{20}{c}}
Y&0\\
0&1
\end{array}} \right)$ and $G_2=\left( {\begin{array}{*{20}{c}}
1&0\\
0&Y
\end{array}} \right)$. They perform as two generators of non-Abelian braid group $B_3$. These matrix characterized group elements belong to group $SO(3)$, thus achievable in our setup. For instance, driving the system along the control loop in Fig.~\ref{fig:4}(a) can realize state transfer as $(\left| \rm{A} \right\rangle,\left| \rm{B} \right\rangle,\left| \rm{C} \right\rangle) \to (\left| \rm{C} \right\rangle,-\left| \rm{A} \right\rangle,-\left| \rm{B} \right\rangle)$. In contrast, the control loop in Fig.~\ref{fig:4}(d) implies state transfer as $(\left| \rm{A} \right\rangle,\left| \rm{B} \right\rangle,\left| \rm{C} \right\rangle) \to (\left| \rm{B} \right\rangle,\left| \rm{C} \right\rangle,\left| \rm{A} \right\rangle)$. {\color{black}We have experimented with these two configurations, and the results in Figs.~\ref{fig:4}(b) and ~\ref{fig:4}(e)  show expected acoustic energy transfers with the efficiency summarized in the upper panel of Figs.~\ref{fig:4}(c) and ~\ref{fig:4}(f).} Another feature of braiding predicts the phase distribution of final states. The onsite energy in our system is unperturbed since all cavities are uniform. Therefore, we successfully observe theoretically predicted phase information of output states shown in {\color{black}the lower panel of} Figs.~\ref{fig:4}(c) and ~\ref{fig:4}(f). The final states, obtained by reversely ordered evolution path, sustain the same dynamic phase and demonstrate the non-commutativity of two braiding operations.

In conclusion, we have experimentally demonstrated the evolution of instantaneous acoustic states via dynamic modulation. {\color{black}The transient system is compact, versatile, and accurate, as verified by the implementation of various logic operations, including the Y gate, Z gate, Hadamard gate, and non-Abelian braiding}. The mechanism relies on producing states with high-quality factors and driving the states along specific paths in parameter space. The evolution can be further accelerated by shortcuts \cite{guery2019shortcuts} or be generalized to a non-adiabatic process \cite{sjoqvist2012non,yan2019experimental}. {\color{black}More importantly, the effective hopping can be nonlinear and the controlled parameter space is not limited to real or continuous values.} These attributes indicate a universal platform that could facilitate the engineering of states beyond the linear Hermitian systems, including universal logic gates \cite{duan2001geometric,kauffman2004braiding}, nonlinear excitation \cite{jurgensen2021quantized,fu2022two}, states evolution at the Riemann surface \cite{nasari2022observation}.

\begin{acknowledgements}
Ze-Guo Chen thanks Haitan Xu, Minghui Lu, Yanfeng Chen for fruitful discussions. The work is supported by the National Key Research and Development Program of China (No. 2022YFA1405000), the Natural Science Foundation of Jiangsu Province, Major Project (No. BK20212004).
\end{acknowledgements}

%\nocite{*}

\bibliographystyle{apsrev4-1} 
\bibliography{Ref_rev}% Produces the bibliography via BibTeX.

\end{document}